\documentclass[lettersize,journal]{IEEEtran}
\usepackage{amsmath,amsfonts}
\usepackage{amsthm} 
\usepackage{algorithmic}
\usepackage{algorithm}
\usepackage{array}
\usepackage{textcomp}
\usepackage{stfloats}
\usepackage{url}
\usepackage{verbatim}
\usepackage{graphicx}
\def\BibTeX{{\rm B\kern-.05em{\sc i\kern-.025em b}\kern-.08em
    T\kern-.1667em\lower.7ex\hbox{E}\kern-.125emX}}
\usepackage{balance}
\usepackage{subcaption}
\usepackage{multirow} 
\usepackage{booktabs}  
\begin{document}

\title{CE-LSLM: Efficient Large-Small Language Model Inference and Communication via Cloud-Edge Collaboration}
\author{{Pengyan Zhu, \and Tingting Yang$^{\ast}$ }
\thanks{$^{\ast}$ Corresponding Author \\
	Pengyan Zhu is with the Navigation College, Dalian Maritime University, Dalian, China (e-mail: winter$\_$zhu@dlmu.edu.cn).\\
	Tingting Yang is with the Peng Cheng Laboratory, Shenzhen, China, and the Navigation College, Dalian Maritime University, Dalian, China (e-mail: yangtt@pcl.ac.cn).\\}}

\markboth{Journal of \LaTeX\ Class Files,~Vol.~18, No.~9, September~2020}%
{How to Use the IEEEtran \LaTeX \ Templates}

\maketitle

\begin{abstract}

Emerging intelligent service scenarios in 6G communication impose stringent requirements for low latency, high reliability, and privacy preservation. Generative large language models (LLMs) are gradually becoming key enablers for the integration of semantic communication and computation. However, due to the limited computational resources of edge devices and the increasing complexity of heterogeneous terminal access, existing centralized inference approaches fail to meet the dual demands of response efficiency and data privacy in edge-side inference tasks. To address these challenges, this paper proposes a novel collaborative inference architecture that integrates cloud-based LLMs with edge-deployed small language models (SLMs), enabling dynamic scheduling and sharing of semantic-level intermediate states, and establishing a unified computation-communication paradigm tailored for 6G networks. Specifically, a key-value (KV) cache reuse mechanism is introduced to enhance the semantic understanding of edge models through contextual guidance from the cloud, while significantly reducing edge-side computational and storage overhead. Furthermore, a cross-node parallel scheduling mechanism is proposed to achieve asynchronous coordination between model state loading and decoding computation, thereby improving edge responsiveness. In addition, we investigate layer alignment and representation compression strategies between heterogeneous models to alleviate the communication burden on the edge. Experimental results demonstrate that the proposed architecture exhibits superior adaptability and scalability in terms of inference latency, system stability, and concurrent processing capacity.

\end{abstract}


\section{Introduction}

\IEEEPARstart{W}{ith} the large-scale commercial deployment of fifth-generation (5G) cellular systems, there is a renewed desire to explore new solutions to solve the real-world problems of relatively small coverage, spectrum resource constraints, and poor device compatibility to improve the existing lifestyles. As the successor to 5G communication technology, sixth-generation (6G) will not just be a wireless network with higher transmission rates and greater network capacity, but a digitally-connected, fully automated and intelligent global communication network capable of delivering diverse network services for different scenarios, including industrial intelligence, smart wearables, autonomous vehicles, and smart healthcare system \cite{ref6G}. Terahertz communication, a cornerstone technology for 6G, leverages electromagnetic waves within the 0.1 THz to 10 THz frequency range to achieve data transmission rates of up to 1 Tbps. This capability supports the high bandwidth and ultra-low latency demands of specialized 6G task scenarios~\cite{white}, \cite{AI-native}, \cite{6G}. However, due to the short wavelength of terahertz waves, combined with their limited penetration and propagation distance, adequate signal coverage requires the deployment of dense small-scale models or edge nodes to establish multi-tier network architectures, encompassing both near-edge and far-edge distributed scenarios. From a practical perspective, while multi-tier network architectures can mitigate signal coverage issues, edge nodes face significant challenges in directly handling complex inference tasks due to constraints in computational power, memory, and bandwidth resources. Furthermore, with the exponential growth in device connectivity, 6G must address not only the insufficiency of computational and transmission resources but also the critical need for robust user privacy protection.      

The emergence of large language models (LLMs), exemplified by ChatGPT, will provide a significant driving force and key support for the development of 6G. Models such as the GPT series and BERT, with billions or even hundreds of billions of parameters, possess high-precision natural language understanding and generation capabilities. They can extract more features and knowledge from vast amounts of data, offering greater potential for solving complex real-world problems. However, due to the immense size of these models, storing and processing their parameters requires substantial memory. For resource-constrained devices, the widespread deployment of LLMs is challenging.

Existing research has improved the training and inference efficiency of LLMs to some extent by restructuring GPU inference cluster architectures and employing advanced hardware-software co-optimization techniquesh~\cite{optimizing},\cite{performance},\cite{deepspeed}. Parallel strategies, such as data parallelism, tensor parallelism, pipeline parallelism, and sequence parallelism, further optimize performance through task division, alleviating memory and computational resource limitations during model inference~\cite{reducing},\cite{gpuclusters}. Additionally, many practical applications involve processing long texts, such as multi-turn conversations, and historical data. Models need to capture long-term dependencies to make accurate predictions. The cloud service framework vLLM draws inspiration from the virtual memory management methods of operating systems to optimize key-value cache management and introduces the concept of prefix sharing~\cite{pagedattention}. By precomputing and storing the attention states of commonly used prompts, studies such as ~\cite{relayattention} and ~\cite{promptcache} further extend the application of prefix sharing, enabling efficient reuse of repetitive prompts.

However, the aforementioned studies are primarily focused on cloud computing. On one hand, the extensive data transmission between the cloud and edge devices results in high latency and bandwidth pressure. On the other hand, privacy protection and data security remain challenging in sensitive task scenarios. These limitations of cloud dependency hinder the development of 6G in edge-oriented applications. To address these issues, small models generated through techniques such as fine-tuning and pruning are deployed on edge devices in various regions to share computational tasks and alleviate the burden on the cloud~\cite{model},~\cite{edgellm}. However, these small models typically require further training, and due to their reduced scale, their output accuracy is often insufficient to fully meet the high-precision inference requirements of complex scenarios.

In modern communication networks, distributed systems are commonly adopted architectures that contribute to reducing the dependence of individual nodes on central servers, thereby enhancing the flexibility and scalability of the network. The concepts of ubiquitous collaboration and interconnectivity have become fundamental principles of 6G. In~\cite{AIGX}, prompt engineering is introduced into the cloud-edge-terminal collaborative architecture to optimize the quality of service (QoS) for generative AI tasks and to establish efficient collaborative connections. Furthermore, in~\cite{CE-CoLLM}, a hierarchical approach is proposed for LLMs, categorizing them based on computational complexity or resource requirements.

To meet the demands of 6G networks for rapid responsiveness, privacy-preserving communication services, and task diversity, we propose CE-LSLM (Cloud-Edge Large-Small Language Model framework), a novel cloud-edge collaborative inference framework that integrates LLM on the cloud with SLMs at the edge. In this architecture, the cloud is primarily responsible for complex semantic tasks such as long-context modeling, while the edge devices handle lightweight local subtasks with lower computational load. To alleviate the computational and memory burdens on the edge, a cross-device cache reuse mechanism is introduced: edge nodes can reuse deep semantic representations generated by the cloud to enhance inference accuracy, and shallow key-value (KV) caches can be shared among local edge devices to enable collaborative inference and load balancing even in disconnected scenarios. Since most inference processes are executed locally, user data remains on the device, thereby further enhancing privacy protection.

A layer matching strategy based on inter-layer distribution similarity is explored and constructed to align the architectural correspondence between the cloud-based LLM and the edge-deployed SLMs. Meanwhile, an attention-head dimensionality reduction strategy is employed to compress key intermediate caches, thereby reducing transmission and storage overhead. Furthermore, a layer-wise pipelined execution mechanism is designed to enable edge devices to preload intermediate states for subsequent layers while computing the current layer, achieving collaborative parallelism between inference computation and communication loading. This design significantly improves the system's response latency and throughput performance.

In summary, the following key contributions have been made:
\begin{itemize}
	
\item We propose a novel collaborative inference architecture integrating cloud-based LLMs and edge-deployed SLMs, offering a unified computation-communication design tailored to the demands of 6G networks.

\item By leveraging both cloud-edge and edge-edge KV cache reuse, combined with a hierarchical and pipelined scheduling strategy, we address the resource constraints of edge devices while meeting complex inference demands in multi-device, multi-task environments.

\item We integrate existing layer matching and attention-head compression techniques into a unified cache-sharing framework tailored to heterogeneous LLM-SLM inference. This enables scalable cache reuse across cloud-edge and peer-edge settings, addressing cross-device compatibility and reducing communication overhead in collaborative inference.

\end{itemize}

We organize the remainder of this article as follows. Section II reviews related work. Section III presents a framework for collaborative cooperation between large and small models in 6G networks. Section IV introduces the system model and communication mechanisms. Section V describes optimization strategies for collaborative inference. Section VI presents experimental evaluations. Section VII concludes the paper. The Appendix provides supplementary materials, such as mathematical proofs.

\begin{figure*}[t]
	\centering
	\includegraphics[width=7in,height=4.5in]{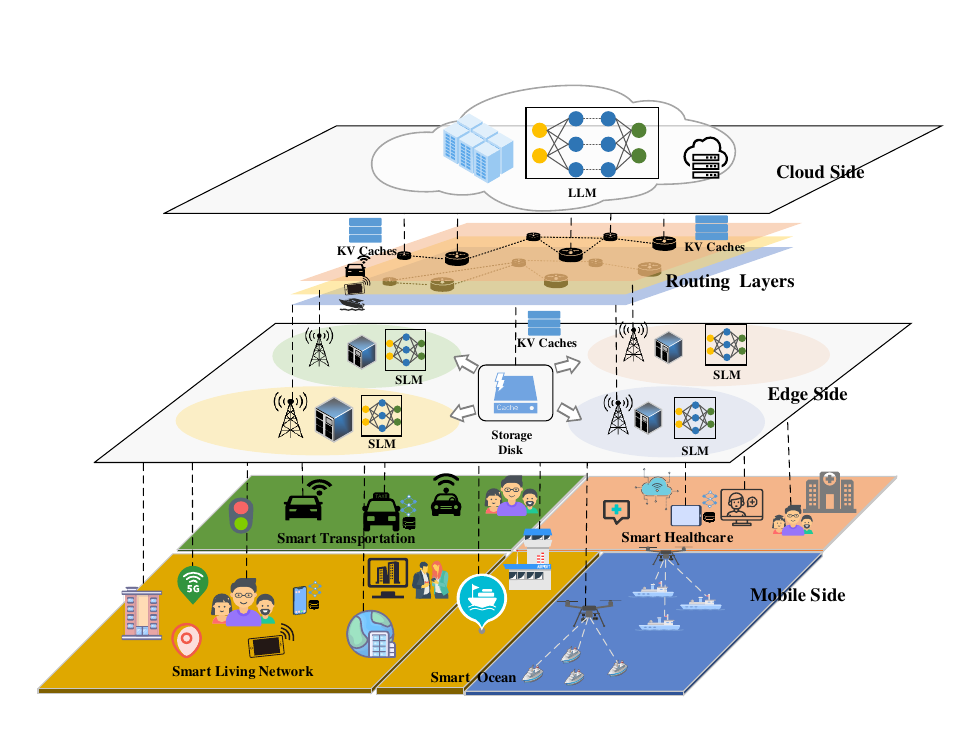}
	\caption{Collaborative Inference Workflow of Cloud LLM and Edge SLMs Across Heterogeneous Scenarios.}
	\label{fig1}
\end{figure*}

\section{RELATED WORK}

Large language models (LLMs), such as GPT-3 and Bloom, typically consist of hundreds of millions to tens of billions of parameters. The resource demands during both training and inference far exceed the memory capacity of a single GPU. To address this challenge, DeepSpeed emphasizes distributed deployment and collaborative mechanisms for multi-GPU inference \cite{deepspeed}\cite{infinite}. Its core component, ZeRO (Zero Redundancy Optimizer), partitions gradients, optimizer states, and other model data during distributed training to eliminate redundancy \cite{zero}. However, these optimizations primarily target the training phase.

During single-query inference, all attention scores for context tokens must be recomputed for each query, requiring frequent access to intermediate results. This issue is exacerbated in long-sequence tasks, where the complexity of attention computation increases from $O\left(n\right)$ to $O\left(n^2\right)$, further amplifying computational and memory access redundancy. vLLM addresses this challenge by focusing on the batched processing of multi-user requests, emphasizing the reuse and sharing of cached data to enhance cache management efficiency on a single GPU~\cite{pagedattention}. Building on this foundation, \cite{RelayAttention} separates system prompts from user prompts, storing intermediate results for system prompts in advance. When computing attention for each batch of user prompts, only a single read of the upstream KV cache is required. This approach partially mitigates the redundant memory access issues associated with processing long-context sequences. However, resource-constrained edge devices face significant challenges in handling the computation of system prompts and the storage of intermediate results. As a result, substantial amounts of client data must still be uploaded to the cloud, leading to high latency and increasing the risk of privacy breaches during data transmission.

Lightweight model deployment strategies have made it feasible to implement large language models (LLMs) on resource-constrained edge devices, alleviating the over-reliance on cloud computing. Model compression techniques, including quantization and pruning, are commonly employed to reduce model size and computational complexity. For instance, \cite{llmpruner} utilizes gradient information to selectively remove non-critical coupled structures, enabling task-agnostic LLM compression, albeit requiring additional training or re-optimization. In contrast to structured pruning methods, works such as \cite{sparse}, \cite{edgellm}, and \cite{cachegen} consider the sensitivity variations across model layers, adapting layer-wise compression strategies accordingly~\cite{kvsharer}. Cache quantization represents another area of research; \cite{adakv} and \cite{discard} propose adaptive KV cache eviction and allocation mechanisms based on attention head characteristics, reducing KV cache size to fit within predefined memory budgets.

Recently, ~\cite{CE-CoLLM} introduced a cloud-edge collaborative inference architecture, in which certain layers of the LLM are deployed on edge devices while others operate in the cloud. This architecture reduces communication overhead through an early exit mechanism. However, in complex tasks such as intelligent healthcare, the limited computational power of edge devices may prevent the generation of high-confidence inference results. As a result, the frequent transmission of intermediate states between the cloud and edge not only significantly increases communication overhead and latency but also undermines privacy protection. Furthermore, in cases of network interruptions or instability, the cloud-based inference in CE-CoLLM cannot be completed, making it impossible to meet the high accuracy requirements of such tasks. In NetGPT~\cite{netgpt}, the collaborative inference paradigm is centered around prompt construction, where lightweight LLMs at the edge are responsible for personalized prompt completion. These edge models are further fine-tuned via low-rank adaptation techniques such as LoRA to enhance their semantic understanding. 

In contrast, our collaborative architecture is better suited to the diverse and complex task requirements of 6G networks. The cloud-based LLM is primarily responsible for handling long-context inference in domain-specific systems, while edge SLMs focus on processing local user requests with lower resource consumption, ensuring that user privacy data does not need to be uploaded to the cloud. By reusing the cloud's KV cache, SLMs are able to achieve more accurate inference results with reduced computational overhead. Even in the case of a cloud-edge network disconnection, the edge devices can still complete inference tasks through collaboration and cache sharing across multiple devices, further enhancing system robustness and inference efficiency.
\begin{figure}[t]
	\centering
	\includegraphics[width=3in,height=4.3in]{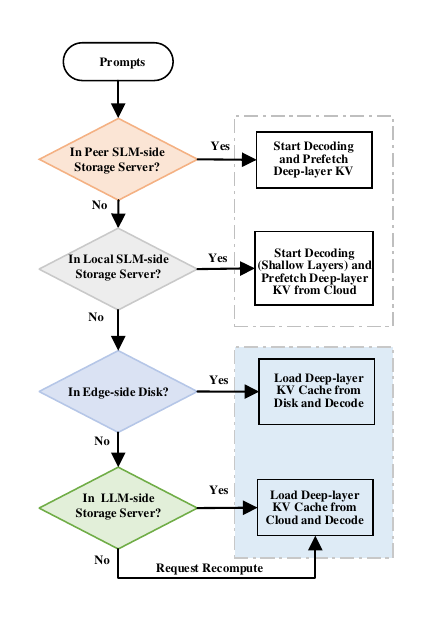}
	\caption{System architecture illustrating multi-level contextual KV cache scheduling and interaction between SLMs and LLM. The SLM first searches for shallow-layer contextual KV caches from peer or local edge storage to enable early-stage decoding. If a match is found, decoding starts while deeper-layer KV blocks—generated by the cloud LLM—are concurrently loaded from local disk or retrieved from cloud storage. This layer-aware, pipelined mechanism enables fast and resilient inference under dynamic connectivity conditions.}
	\label{fig2}
\end{figure}
\begin{figure*}[t]
	\centering
	\includegraphics[width=7in,height=3.8in]{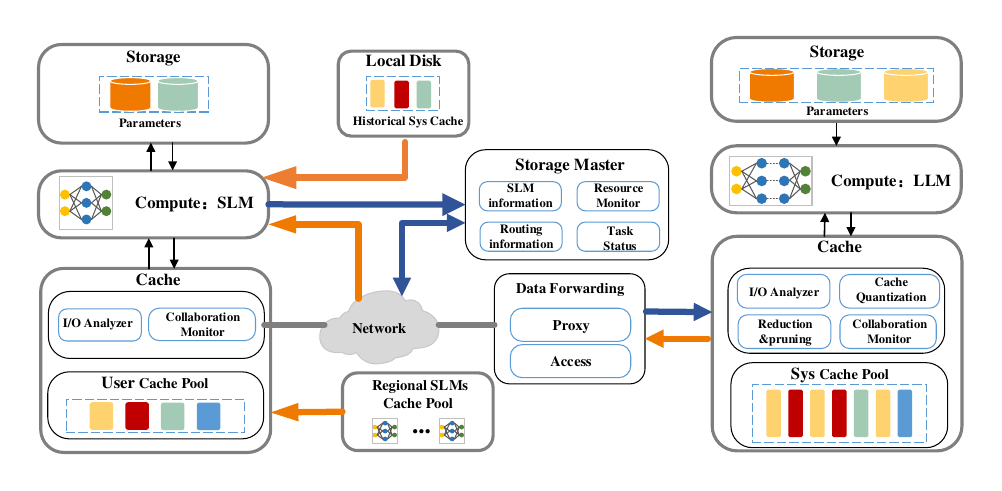}
	\caption{Schematic of Cache and Storage System Interaction Between LLM and SLM. From left to right, the diagram illustrates the edge device storage and caching module, the data forwarding layer, and the cloud storage and caching module. During the execution of complex inference tasks, the edge SLM requests and downloads the optimized contextual KV cache from the cloud-based LLM.}
	\label{fig3}
\end{figure*}
\section{Design and Implementation of a Cloud-Edge Collaborative Inference Architecture}

In this section, the fundamental components and overall workflow of the cloud-edge collaborative inference architecture for LLMs and SLMs are first introduced in detail. Subsequently, the dynamic caching strategy is presented, which serves as a key enabler for the efficient allocation, transmission, and utilization of KV caches, acting as a critical bridge for collaborative inference.

\subsection{Framework Overview}

In numerous real-world application scenarios, such as intelligent transportation, healthcare, the Internet of Things (IoT), and smart ocean systems, domain-specific task frameworks rely heavily on the correlation analysis of long-term historical data and global inference. Cloud servers are equipped with high-bandwidth GPUs and large-scale memory resources, enabling efficient support for large-scale language models (LLMs) in terms of both parameter loading and long-sequence Key-Value (KV) cache computation and storage. As core intelligent components, LLMs possess powerful capabilities in contextual understanding and generation, allowing them to extract high-level semantic features from input sequences, including long-range dependencies, cross-sentence structures, and implicit task intentions. However, edge devices are significantly constrained in terms of computational power, memory, and storage, making it infeasible to load the full-scale LLM or manage the extensive KV cache required for long-context inference. As a result, Smaller-scale Language Models (SLMs) deployed at the edge are typically responsible for modeling the lower-layer KV caches of system prompts, generating local interactions involving privacy-sensitive user data, and responding quickly to user queries—thereby striking a balance between data security and service latency optimization. Fig.~\ref{fig1} illustrates the basic components and information flow of the architecture.

The cloud-based LLM generates the key-value (KV) cache for the background information (i.e., system prompts) through the attention mechanism, providing contextual support for the inference tasks of SLMs. Based on the model's structural parameters, the generated KV cache is subjected to hierarchical pruning and dimensionality reduction, and it is stored in the cloud service or distributed to the local disk of edge devices to support multi-device sharing and reuse.

The inference process of the edge-side SLM is divided into two stages: on one hand, the shallow-layer KV caches of the system prompt are computed locally, while deep-layer KV caches are progressively loaded from the cloud or local disk. On the other hand, once the KV cache for a given layer of the system prompt is prepared, the corresponding Q, K, and V computations for the user prompt are immediately executed. Through reserved encoding positions, the KV caches from both local and remote sources are concatenated into a complete contextual KV sequence, which is then used to generate the attention output for that layer. As the deep-layer cache loading is completed, the computation concludes. The final inference result is then returned to the user. The cache loading process is illustrated in Fig.~\ref{fig2}.

The design of this architecture is based on several key considerations. First, loading deep-layer KV caches helps to enhance the semantic information of edge outputs and alleviate the computational burden on the SLM, while shallow-layer KV caches, being more sensitive to generative behaviors, need to be generated locally by the edge SLM. Second, local computation is typically faster than remote KV loading, which helps to accelerate decoding initiation. Meanwhile, edge devices do not require the preloading of all cloud-based KV caches, significantly alleviating storage pressure ~\cite{computeorload}. Therefore, a layer-wise pipelined decoding mechanism is adopted, enabling parallel execution of computation and cache loading, as well as coordinated processing of system prompts and user requests, which effectively accelerates user-prompt decoding and enhances overall inference performance.

\subsection{Dynamic Cache Management System}
In traditional communication systems, the core value lies in reliably transmitting raw or processed data from the sender to the receiver while ensuring data integrity and consistency. However, KV caches, as intermediate states in model inference, encapsulate substantial semantic information. They are not merely data but also directly usable inference units. This transforms communication networks from simple data forwarding systems into collaborative communication frameworks that support intelligent inference and generation. Within this framework, the cache management system becomes a critical component, responsible for the dynamic allocation and optimization of KV caches to ensure efficient collaboration between the cloud and edge models. Due to the differing functional roles of the cloud and edge in collaborative inference, the design of their cache management systems also varies, as illustrated in Fig.~\ref{fig3}.

The Storage Server is primarily responsible for the long-term storage of core data such as model weight parameters, whereas the Cache Server is utilized for temporarily storing Key-Value (KV) caches of prompts. Its submodules include the Collaboration Monitor, which continuously monitors edge requests and system coordination status to ensure efficient cache allocation, and the I/O Analyzer, which analyzes the input/output access patterns of cached data. In addition, a cache optimization module is employed to dynamically adjust the quantization precision and prune redundant layers or dimensions in order to accommodate the inference requirements of both LLMs and SLMs. The storage and caching modules on the edge are designed similarly to those on the cloud. However, in the edge-side Local Cache, a historical cache module is deployed to store system prompt KV caches periodically downloaded from the cloud. This mechanism allows the cached content to serve as backup context during network disconnection, thereby maintaining inference continuity.

The communication system consists of two modules: Storage Master and Data Forwarding. The Storage Master provides information such as the structure of the edge SLM and network bandwidth, which is then synchronized with the Proxy module. The Proxy module is responsible for making transmission path decisions, selecting either direct (point-to-point) transmission or routing data to the cloud, based on the network status. In case of network anomalies, the Proxy retrieves the context cache from the edge device’s disk, utilizing the local cache to complete the inference. The Access module is responsible for the actual data transmission tasks~\cite{osca}.

\section{SYSTEM MODEL AND OPTIMIZE CLOUD-EDGE INFERENCE}\label{sec3}

This section focuses on the key computational mechanisms of cloud-edge collaboration and an analysis of communication performance. First, the layer-by-layer attention computation and KV cache reuse mechanisms, which form the foundational basis of cloud-edge collaborative inference, are described in detail. Subsequently, a communication performance model is proposed to quantify the time delay resulting from data transmission during collaboration. This model provides a theoretical basis for developing efficient transmission strategies in 6G networks.

\subsection{System Model} \label{infere}

A causal encoder adopts a decoder-only Transformer architecture and is widely used in autoregressive generation tasks, such as text generation and language modeling.
In this architecture, the decoder predicts and generates one token at each time step, relying solely on previously generated tokens as contextual information, with no access to future positions. Leveraging this property, the representations generated from the system prompt can be reused as contextual inputs for user prompt inference. By appropriately partitioning the inference responsibilities of system and user prompts, the collaborative inference framework enables upstream–downstream decoupling and computational load balancing. 

We consider a single attention head where the system prompt has a length of $s$, the calculations of the query, key, and value for the $l$-th layer, for all $l \in \left\{1,2,\dots, N\right\}$, are expressed as follows :
\begin{equation}
	\begin{aligned}
	&q^{\left ( l \right ) }_{ctx,i}=x^{\left ( l \right ) }_{ctx,i}\cdot W_{Q}^{\left ( l \right ) }\\
	&k^{\left ( l \right ) }_{ctx,i}=x^{\left ( l \right )}_{ctx,i}\cdot W_{K}^{\left ( l \right )} \\
	&v^{\left ( l \right ) }_{ctx,i}=x^{\left ( l \right )}_{ctx,i}\cdot W_{V}^{\left ( l \right )}\\
	\end{aligned}
\end{equation} 
where $i\in\left\{1,2,…,s\right\}$ indexes the position of tokens within the system prompt sequence, and $x_{ctx}^{(0)}=\gamma \cdot r_{ctx} + b$ denotes the transformed input features $r_{ctx}$ after applying linear transformations (e.g., scaling and bias). $W_{Q}$, $W_{K}$ and $W_{V}$represent the query, key and value weight matrices of each attention head in the cloud model.

Then, the attention output can be defined as:
\begin{equation}
	\begin{aligned}
		o_{s}^{\left(l\right)}&=Attention\left(q_{s}^{\left ( l \right )},\left \{ k^{\left ( l \right ) }_{i}\right \}_{i=1}^{s}, \left \{ v^{\left ( l \right ) }_{i} \right \}_{i=1}^{s} \right)\\
		&=\sum_{i=1}^{s} \frac{exp\left ( q_{s}^{\left ( l \right ) } \left ( k_{i}^{\left ( l \right )} \right ) ^{T}\right ) }{ \sigma^{\left ( l \right ) }_{1\rightarrow s }}v^{\left ( l \right ) }_{i} 
	\end{aligned}
\end{equation}
where $\sigma^{\left ( l \right ) }_{1\rightarrow s }= {\textstyle \sum_{i=1}^{s}} exp\left ( q_{s}^{\left ( l \right ) } \left ( k_{i}^{\left ( l \right )} \right ) ^{T}\right )$ is used to normalize the attention weights. the output $o_{s}^{\left(l\right)}$ will serve as the input to the next layer, continuing the attention calculation until the entire inference process is completed. 

The cloud model stores the intermediate results $k_{ctx}^{\left(l\right)}$ and $v_{ctx}^{\left(l\right)}$ for each layer, which are made available for SLMs to download and utilize. In addition, within a local region, multiple SLMs can collaboratively compute or share these contextual representations, thereby reducing redundant computation and improving memory efficiency. Meanwhile, the output serves as the input to the next layer $x^{\left ( l+1 \right ) }_{ctx,s}=o_{s}^{\left(l\right)}$, enabling the continuation of attention computation until the inference process is completed.

\begin{figure}[t]
	\centering
	\includegraphics[width=3.6in,height=3in]{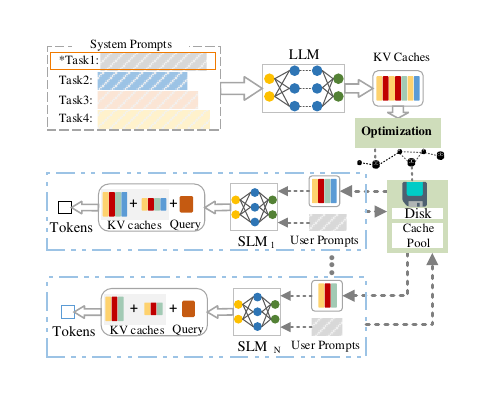}
	\caption{ Illustration of Network-Resilient LLM Inference via Cache Reuse and Token Generation in Edge SLMs.}
	\label{fig4}
\end{figure}

Based on system prompt processing, the edge-side SLM is responsible for the localized computation and caching of the user prompt and the partially generated response tokens. Let the user prompt have a length of $u$, and assume that $t-1$ response tokens have already been generated. Then, the tokens from position $s+1$ to  $L=s+u+t-1$ participate in the current attention computation. The corresponding query, key, and value representations are computed as follows:
\begin{equation}
	\begin{aligned}
		&q^{\left ( l \right ) }_{user,j}=x^{\left ( l \right ) }_{user,j}\cdot W_{Q,SLM}^{\left ( l \right ) }\\
		&k^{\left ( l \right ) }_{user,j}=x^{\left ( l \right )}_{user,j}\cdot W_{K,SLM}^{\left ( l \right )}  \ \\
		&v^{\left ( l \right ) }_{user,j}=x^{\left ( l \right )}_{user,j}\cdot W_{V,SLM}^{\left ( l \right )}\\
	\end{aligned}
\end{equation} 

The attention output can be defined as:
\begin{equation}
	\begin{aligned}
		o_{L}^{\left(l\right)}&=Attention\left(q_{L}^{\left ( l \right )},\left \{ k^{\left ( l \right ) }_{j}\right \}_{j=s+1}^{L}, \left \{ v^{\left ( l \right ) }_{j} \right \}_{j=s+1}^{L} \right)\\
		&=\sum_{j=s+1}^{L} \frac{exp\left ( q_{L}^{\left ( l \right ) } \left ( k_{j}^{\left ( l \right )} \right ) ^{T}\right ) }{ \sigma^{\left ( l \right ) }_{s+1\rightarrow L }}v^{\left ( l \right ) }_{j} 
	\end{aligned}
\end{equation}
where $\sigma^{\left ( l \right ) }_{s+1\rightarrow L }= {\textstyle \sum_{j=s+1}^{L}} exp\left ( q_{L}^{\left ( l \right ) } \left ( k_{j}^{\left ( l \right )} \right ) ^{T}\right )$ is used to normalize the attention weights.

In summary, for a batch of inference requests originating from user devices, the edge server coordinates the collaborative inference process between the cloud-based LLM and the local SLMs. At generation step $t=L+1$, the attention output $o_t$ of SLM is defined as follows:
\begin{equation}\label{ot}
	\begin{aligned}
		o_{ t,SLM }=&\alpha_{ctx}^{t}Atttention\left ( q_{t}, \left \{  k_{i}\right \}_{i=1}^{s}, \left \{ v_{i} \right \}_{i=1}^{s}\right )+\\
		&\alpha_{user}^{t}Atttention\left ( q_{t}, \left \{  k_{j}\right \}_{j=s+1}^{L}, \left \{ v_{j} \right \}_{j=s+1}^{L}\right ) \\
		s.t. \ &\alpha_{ctx}^{t} =\frac{\sigma^{\left ( l \right ) } _{1\rightarrow s }}{\sigma^{\left ( l \right ) } _{1\rightarrow L }},\ \alpha_{user}^{t}=\frac{\sigma^{\left ( l \right ) } _{s+1\rightarrow L }}{\sigma^{\left ( l \right ) } _{1\rightarrow L }}\\
		&\alpha_{ctx}^{t}+\alpha_{usr}^{t} =1
	\end{aligned}
\end{equation}

This hierarchical collaborative inference architecture offers significant advantages. On one hand, resource-constrained edge models can avoid repeatedly computing long system prompts, thereby improving the throughput for handling user requests. On the other hand, the KV cache of the system prompt can be reused across multiple SLMs instances, significantly enhancing overall resource efficiency. Fig.~\ref{fig4} illustrates the workflow of a specific task (e.g., Task 1), in which multiple edge SLMs share the same system prompt KV cache and integrate it with user prompt caches to generate personalized outputs.

\subsection{Communication Performance}
In cloud–edge collaborative inference, communication overhead caused by task partitioning is as critical as computational load, and this trade-off becomes more prominent under 6G network conditions.
The inference latency of LLMs is commonly estimated based on the total number of  $FLOPs$ and the volume of $I/O$ bytes transferred. Hence, we have:
\begin{equation}
	\begin{aligned}
	T_{com_{C}} &=\sum_{l=0}^{N-1} t_{comp_{C}}^{\left ( l \right )} =\sum_{l=0}^{N-1} t_{FLOPs}^{\left ( l \right ) }+t_{I/O}^{\left ( l \right ) }+ t_{decode}^{\left ( l \right )}\\
	\end{aligned}
\end{equation}
where $N$ denotes the total number of layers in LLM, $ t_{FLOPs}^{\left ( l \right ) }=\frac{FLOPs}{TFLOPs/s}$ and $t_{I/O}^{\left ( l \right ) }=\frac{I/O\ bytes}{B_{m}}$ represent the floating-point computation time and the data input/output time for the $i$-th layer on the edge device, respectively. $B_{m}$ is Memory Bandwidth.

Similarly, the inference time of the SLM is expressed as:
\begin{equation}
	\begin{aligned}
		T_{com_{E}} &=\sum_{l=0}^{M-1} t_{comp_{E}}^{\left ( l \right )} =\sum_{l=0}^{M-1} t_{FLOPs}^{\left ( l \right ) }+t_{I/O}^{\left ( l \right ) }+ t_{decode}^{\left ( l \right )}\\
	\end{aligned}
\end{equation}
where $M$ denotes the total number of layers in SLM.

The transmission latency between cloud–edge and edge–edge devices is affected by the available bandwidth and can be expressed as:
\begin{equation}
	\begin{aligned}
		T_{comm} &=\sum_{l=0}^{N-1} t_{comm}^{\left ( l \right )} =\sum_{l=0}^{N-1} \frac{D^{\left ( l \right )}}{B_{t}^{\left ( l \right )}}
	\end{aligned}
\end{equation}
where $D$ denotes the size of kv cache, and $B_{t}$ represents the network bandwidth at the current time step.

\begin{figure*}[t]
	\centering
	\includegraphics[width=7in,height=2.8in]{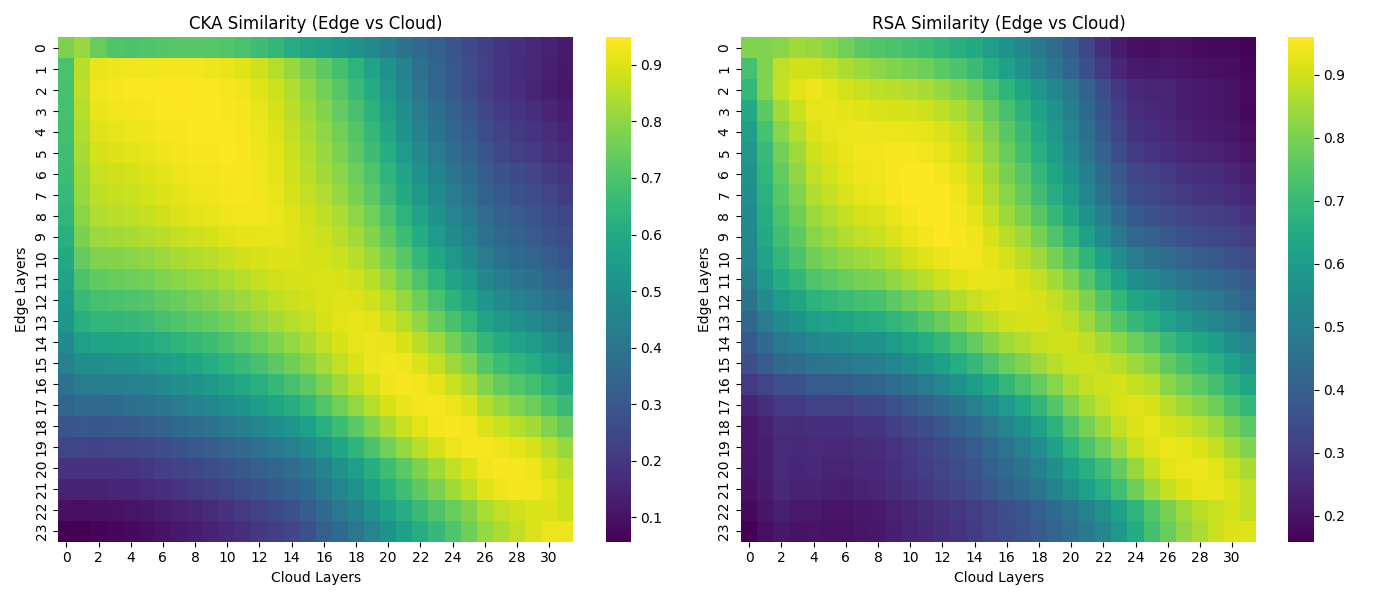}
	\caption{
		Heatmaps of structural similarity between edge and cloud model layers.
		(\textbf{Left}) Centered Kernel Alignment (CKA) similarity scores.
		(\textbf{Right}) Representational Similarity Analysis (RSA) scores.
		Brighter areas indicate higher similarity.
	}
	\label{fig5}
\end{figure*}

The total inference time can thus be calculated by  
\begin{equation}
	\begin{aligned}
		T_{total} =T_{com_{C}} +T_{comm}+T_{com_{E}} 
	\end{aligned}
\end{equation}

The ultra-low latency requirements of 6G networks emphasize the need for real-time responsiveness. Through the cloud-edge collaborative inference architecture, we can achieve fast responses to a large number of user requests while ensuring the privacy protection of local data. Therefore, the problem of minimizing inference latency can be formulated as follows:
\begin{equation}\label{key11}
	\begin{aligned}
		min\ & T_{total}  \\
		s. t. \  &\max \left ( \left \{  Req_{l}\ | \ 1\le l\le N \right \}  \right ) \le Mem_{E}\\
		&\max \left\{  O_{l}\ |\ 1\le l\le N \right\}\le B_{l} \times T_{max}
	\end{aligned}
\end{equation}
where $Req_{l}$ represents the memory required for the $l$-th layer's KV caches in cloud, and $Mem_{E}$ represents the available memory in the edge SLM. Additionally, $o_{l}$ represents the amount of data to be transmitted, $B_{l}$ denotes the transmission bandwidth, and $T_{max}$ is the maximum tolerable transmission delay. 

\section{Optimize Communication Transmission latency} \label{comm}

This section presents a set of techniques aimed at alleviating the computational and communication bottlenecks in collaborative inference. By introducing layer-wise matching and attention head dimensionality reduction, cache reuse between heterogeneous models becomes feasible, allowing the cloud model to enhance the contextual understanding capability of edge models without increasing their computational burden. Furthermore, by optimizing the scheduling of computation and cache loading, the framework enables joint acceleration of computation and transmission, thereby significantly reducing inference latency and improving overall system throughput.

\subsection{Layer-wise Similarity Matching for Cloud–Edge Models}

As system prompts vary across tasks, they influence the activation patterns of LLM, leading to differences in the statistical distributions of KV caches across layers. Prior studies have shown that KV caches from structurally similar layers can be interchangeable without significantly affecting the final hidden state outputs ~\cite{cachegen}. Motivated by this, we propose an exploratory layer-wise matching strategy based on representational similarity, aiming to facilitate KV cache reuse in cloud–edge collaborative inference.Specifically, we select $n$ out of  $N$ layers of LLM and transmitted to SLM. The hidden features of the remaining layers in the LLM to provide long-context KV caches for the SLMs, while the remaining $M-n$ layers are computed locally by the current SLM or by neighboring SLMs of the same type, thereby enabling adaptive cache sharing under dynamic and heterogeneous tasks.

Specifically, when comparing the spatial structures of two layers, directly comparing their raw activation values is often affected by factors such as permutation, scale, and transformation. To address this, representational similarity matrices (RSMs)~\cite{similarity} are employed to capture the structural similarity across representations by computing the similarity between each instance ii and all other instances. The mathematical formulation is given as~\cite{similarity}:
\begin{equation}\label{kenel_matrix}
	S_{ij}^{\left(l\right)}:=s(O_{i}^{\left(l\right)},O_{j}^{\left(l\right)})
\end{equation}
Here, $O_{i}$ and $O_{j}$ denote the $i$-th and $j$-th rows of the output matrix $O\in \mathbb{R}^{N\times D} $ from the $l-th$ layer, where $N$ s the number of input samples and $D$ is the hidden size of the current layer. $S \in \mathbb{R}^{N\times N}$ is a symmetric matrix, and $s \left(\cdot, \cdot \right)$ represents a generic similarity function.
In the following, we jointly evaluate the similarity between heterogeneous models from both spatial and behavioral perspectives.

Compared with other similarity measures, Centered Kernel Alignment (CKA) possesses several desirable invariance properties, with their theoretical proof provided in Appendix~\ref{appendix:CKA}. Therefore, CKA is more suitable for analyzing the intrinsic structural similarity between layer-wise representations in neural networks, rather than superficial differences such as numerical scaling or neuron ordering.

HSIC is used to quantify the statistical dependence between two kernel matrices, where a larger value indicates greater structural similarity. Given two layer representation matrices $R$ and $R^{'} \in \mathbb{R}^{N\times D}$, the computation is expressed as~\cite{similarity}:
\begin{equation} \label{HSIC}
	\begin{aligned}
	HSIC\left ( S_{e}^{\left(l_{e}\right)},S_{c}^{\left(l_{c}\right)}\right ) &=\frac{1}{\left ( N-1 \right )^{2} } tr\left ( H_{N}S_{e}^{\left(l_{e}\right)}H_{N}S_{c}^{\left(l_{c}\right)} \right ) \\
	s.t. \ \ H_{N}&=I_{N}-\frac{1}{N}1_{N}1_{N}^{T}
    \end{aligned} 
\end{equation}
where $S_{c}^{\left(l_{c}\right)}=O_{c}^{\left(l_{c}\right)}O_{c}^{T}$  denotes the kernel similarity matrix of the $l_{c}-th$ layer in the cloud model, computed from its output representation matrix $O_{c}^{\left(l_{c}\right)}$. $S_{e}^{\left(l^{\left(l_{e}\right)}\right)}=RR'^{T}$ denotes the kernel similarity matrix of the edge model at layer $l_{e}$. $H_{N}$ is the centering matrix, which is used to center both kernel matrices, enabling the evaluation of their structural alignment.

Normalization is a critical step that endows the similarity measure with a set of desirable mathematical properties. Specifically, it is expressed as~\cite{similarity}:
{\small
\begin{equation} \label{CKA}
	\begin{aligned}
	CKA\left (  O_{e}^{\left(l_{e}\right)},O_{c}^{\left(l_{c}\right)}\right ) =\frac{HSIC\left ( S_{e}^{\left(l_{e}\right)},S_{c}^{\left(l_{c}\right)}\right ) }{\sqrt{ HSIC\left ( S_{e}^{\left(l_{e}\right)},S_{e}^{\left(l_{e}\right)}\right )\cdot HSIC\left ( S_{c}^{\left(l_{c}\right)},S_{c}^{\left(l_{c}\right)} \right )   } } 
	\end{aligned}
\end{equation}
}
where $ m_{CKA} \left (  O_{e}^{\left(l_{e}\right)},O_{c}^{\left(l_{c}\right)}\right )\in \left [ 0,1 \right ]$~\cite{DBLP}.

We also introduce a constraint on attention behavior consistency.
Specifically, in autoregressive language models, a strict lower triangular attention mask is applied at each layer to ensure that each token can only attend to preceding positions.
Cosine similarity is first used to compute the pairwise similarity between all samples within the same layer ll of the SLM, as defined in~\cite{DBLP}, and is given by:
\begin{equation}
	\begin{aligned}
		S_{RSA,e}^{\left(l_{e}\right)}=\frac{O_{e,i}^{\left(l_{e}\right)}\cdot O_{e,j}^{\left(l_{e}\right)}}{\left \|O_{e,i}^{\left(l_{e}\right)}  \right \| \left \|O_{e,j}^{\left(l_{e}\right)}  \right \| }
	\end{aligned}
\end{equation}
Similarly, $S_{RSA,c}^{\left(l_{c}\right)}$ can be obtained.

Subsequently, the lower triangular part of each similarity matrix is extracted and flattened into vectors $s_{RSA,e}^{\left(l_{e}\right)}$ and $s_{RSA,c}^{\left(l_{c}\right)}$. The Pearson correlation coefficient between these two vectors is then computed to quantify their alignment. So,
\begin{equation}
	\begin{aligned}
		RSA\left (O_{e}^{\left(l_{e}\right)},O_{c}^{\left(l_{c}\right)} \right )  =\text{Corr}\left(s_{RSA,e}^{\left(l_{e}\right)}, s_{RSA,c}^{\left(l_{c}\right)}\right)
	\end{aligned}
\end{equation}

Given predefined similarity thresholds $\theta_{CKA}$ and $\theta_{RSA}$, we compute the similarity between a target layer $l_{e}\in \left\{1,2,…,n\right\}$ in the edge model and all candidate layers  $l_{c}\in \left\{1,2,…,N\right\}$ in the cloud model, in order to evaluate their structural and semantic alignment. Subsequently, the most structurally aligned cloud layer  $l^{*}_{c}$ with respect to $l_{e}$ is selected such that it satisfies the following threshold condition:
\begin{equation}
	\begin{aligned}
		l^{*}_{c} =\arg\max s\left (O_{e}^{\left(l_{e}\right)},O_{c}^{\left(l_{c}\right)} \right ) \\
	s.t. \ \  m_{CKA}\left (O_{e}^{\left(l_{e}\right)},O_{c}^{\left(l_{c}\right)} \right )\ge \theta_{CKA}\\
		RSA\left (O_{e}^{\left(l_{e}\right)},O_{c}^{\left(l_{c}\right)} \right )\ge \theta_{RSA}
	\end{aligned}
\end{equation}
where $s(\cdot, \cdot)$ denotes a general similarity function between the layer representations.

During inference, shallow layers tend to capture fundamental linguistic features such as grammar and syntax. These features are critical for downstream semantic understanding and are thus more sensitive to information loss. In contrast, deeper layers predominantly encode semantic details and abstract contextual features, which are relatively less stringent in their error tolerance. As shown in Fig.5, we visualize the structural similarity matrices between edge and cloud models using both CKA and RSA metrics. Brighter regions indicate higher layer-wise similarity. Notably, the similarity maps exhibit a clear diagonal trend, suggesting a strong layer-wise alignment across the two heterogeneous models. Based on these observations, we define our layer sharing strategy: among candidate pairs that satisfy the similarity threshold, we prefer to select shallower layers for sharing to preserve core semantic representations. The final set of KV cache sharing layers is defined as $L_{Shared}=\left \{  L_{1},L_{2},\cdots ,L_{n}\right \} $. It is worth noting that if the SLMs deployed within a region share the same architecture, shallow-layer KV caches can be directly shared across nodes without the need for structural alignment or layer matching.

\subsection{Dimensionality Reduction via think}

To further reduce the computational and transmission overhead of attention heads during collaborative inference, we introduce a dimensionality reduction strategy inspired by the pruning objective proposed in \cite{think}. Specifically, for each attention head ii, we aim to select a subset of channels that preserve the main interaction patterns between queries and keys. The pruning objective is formulated as~\cite{think}:
\begin{equation}
	\begin{aligned}
		\min_{S}& \left \| Q_{i}K_{i}^{T}-Q_{i}S\left ( k_{i}S \right )  ^{T}\right \|_{F}\\
		s.t.\ \ &trace\left ( S \right )  =\left \lfloor \left ( 1-\lambda  \right )D  \right \rfloor 
	\end{aligned}
\end{equation}
where $Q_{i}\in\mathbb{R} ^{S\times D}$ and $K_{i}\in\mathbb{R}^{S\times D}$ denote the query and key matrices of the $i-th$ attention head, with sequence length $S$ and head dimension $D$. $S\in\left \{ 0,1 \right \}^{D\times D}$ is a binary diagonal matrix, indicating whether a given channel (dimension) is retained 1 or pruned 0. $\lambda \in\left [ 0,1 \right ]$ controls the pruning ratio, such that only $\left ( 1-\lambda  \right ) D$ channels are preserved, and $\left \| \cdot  \right \|_{F}$ is the Frobenius norm.

When the dimensionality of KV cache vectors is reduced from $d_{c}$ to $d_{e}$, we adopt the estimation formula proposed in prior work \cite{survey} to quantify the reduction in computational cost and transmission overhead. Specifically, the reductions in computational FLOPs and I/O transmission volume are expressed as:
\begin{equation}
	\begin{aligned}
		\triangle _{FLOPs}&=L\cdot 8bmk\left (d_{c}-d_{e}  \right )\\ 
		\triangle _{I/O}&=L\cdot \left(4bmk\left (d_{c}-d_{e}  \right ) +4bk\left (d_{c}-d_{e}  \right ) \right)
	\end{aligned}
\end{equation}
where $b, m, k, d$, and $h = k \cdot d$ represent the batch size, sequence length, number of attention heads, hidden size of each head, and overall hidden size, respectively.

Assuming model parameters$b=1,m=1024,k=32,d_{c}=80,d_{e}=64$ ,and $L=32$, the changes in computational cost and data transmission requirements are calculated as $\triangle _{FLOPs}=134217728 FLOPs$, and $\triangle _{I/O}=66.9MB$. Under the conditions of an edge device with a floating-point computational capacity of 100 GFLOPs and a communication bandwidth of 10 Mbps, dimensionality reduction reduces communication latency by 6.69 seconds and inference time by approximately 1.34 milliseconds. These results demonstrate that optimizing the KV cache significantly reduces communication costs between the cloud and edge, as well as the computational burden of inference on edge devices.

\subsection{Communication total optimization}
As a key enabling technology in 6G, terahertz communication offers ultra-high bandwidth for edge devices. However, its extremely short wavelength and limited penetration capacity significantly constrain the effective transmission range, making multi-hop relaying infeasible in many scenarios. As a result, direct transmission between the cloud and edge or among edge devices is preferred to minimize the latency caused by link hopping. 
\begin{figure}[t]
	\centering
	\includegraphics[width=3.8in,height=3in]{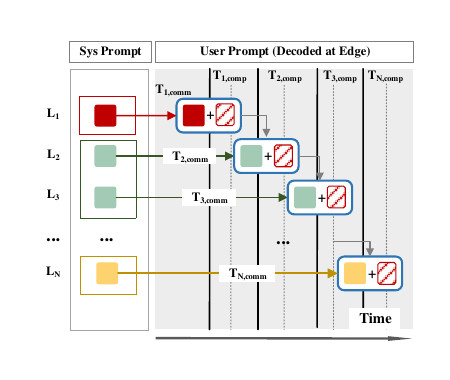}
	\caption{Pipeline Parallel Optimization Strategy. The horizontal axis represents time, while the vertical axis depicts the hierarchical structure of the model.}
	\label{fig6}
\end{figure}

According to the layer-wise similarity analysis (see Eq. 16), the deeper layers of the edge SLMs are generally more structurally aligned with those of the cloud LLM. As a result, the edge SLM prioritizes reusing deep-layer contextual KV caches generated by the cloud, with dimensionality reduction applied when necessary to enhance its capability for long-range context understanding. In addition, CE-LSLM also supports cache sharing among regional edge devices. When one node completes the KV cache computation for the shallow layers of the system prompt ahead of others, the results can be transmitted over the local interconnect to accelerate the user prompt processing of other local SLMs. Let $M$ denote the total number of layers in SLM, and $n$ represent the number of layers whose contextual caches are reused from LLM. We define the cache source function $src\left(l\right)$ for the $l-th$ layer as:
\begin{equation}
	\begin{aligned}
		src\left ( l \right )= \begin{cases}
			& arg \min_{\eta\in \left \{ local,peer \right \} } cost_{\eta}\left ( l \right ) ,\ \ l< n \\
			& cloud ,\ \    \        l\ge n
		\end{cases} 
	\end{aligned}
\end{equation}
where $cost_{\eta}\left ( l \right )$ represents the cost of obtaining the cache for layer $l$ from source $\eta$.

To alleviate inference latency caused by sequential execution, a pipelined scheduling mechanism is introduced to enable the parallel coordination between attention computation and contextual KV cache transmission. In traditional processing paradigms, the edge SLM must first complete the KV cache computation for the entire context (e.g., system prompt) before user prompt decoding can be initiated. In contrast, under the proposed strategy, the edge SLM begins computing the contextual KV caches layer by layer from the shallow layers (layer $1$ to $M - n$), which are either computed locally or reused from nearby peers. Meanwhile, the deep-layer caches (layer $M-n+1$ to $M$) are concurrently loaded from the cloud. These two computation and transmission paths converge at the intermediate layer, allowing user prompt inference to begin even before all context caches are fully prepared, thereby significantly reducing the overall response latency. This parallel process can be formulated as follows:
\begin{equation} \label{key7}
	T_{pip}^{\left(l\right)}=\begin{cases}
		& \max \left\{ t_{comm}^{\left(l\right)}\left ( src\left ( l \right ) \right ) ,	t_{comp}^{\left(l-1\right)}\right\}, \ \ l\in\left[1, M-n\right]\\
		& \max \left\{ t_{comm}^{\left(l\right)}\left ( cloud \right ) ,	t_{comp}^{\left(l-1\right)}\right\}, \ \ l\in\left[M-n+1, M\right]
	\end{cases}
\end{equation}
where $t_{comm}^{\left(l\right)}\left(\cdot\right) $ denotes the communication time required to load the KV cache for layer $l$, and $t_{comp}^{\left(l-1\right)}$ represents the computation time of the previous layer.

Fig.~\ref{fig6} visually illustrates the parallel overlapping process of computation and transmission times across layers during the pipelined procedure. The solid vertical lines represent the time consumed for transmitting each layer's cloud-side cache to the edge device, while the gray dashed lines indicate the time required for the edge device to compute the user prompt. Since the computation of the previous layer $l-1$ and the transmission of the next layer $l$ are performed in parallel, the overlapping portion of computation time is covered, and only the larger value between the computation and transmission time is retained.

\section{Experiments}
This section presents a comprehensive evaluation of the proposed CE-LSLM collaborative inference framework. We compare three LLM deployment strategies: cloud-only, edge-only, and cloud-edge collaboration, focusing on key performance metrics such as text generation fidelity, inference latency. 

\subsection{Experimental Setup}

Opt-2-6.7B is deployed as the cloud-based LLM, responsible for long-context inference, while Opt-2-1.3B serves as the edge-side SLM, performing local inference. OPT, an open-source large language model developed by Meta, adopts a decoder-only causal architecture and supports up to 2048 positional tokens, making it well-suited for autoregressive natural language processing tasks. We extract both short- and long-sequence experimental data from the ShareGPT$\_$V3 dataset, which contains extensive real-world user interactions with ChatGPT across various scenarios. Additionally, we utilize the XSum dataset, which comprises BBC articles and serves as a benchmark for evaluating the text generation quality of LLMs. Regarding hardware selection, we employ the NVIDIA A800, a high-performance cloud computing device commonly used in both cloud and edge computing environments. The detailed hardware specifications are presented in Table ~\ref{tab1}.

While vLLM-based inference services exhibit strong asynchronous scheduling and high throughput in practical deployments, our lab environment employs shared GPU resources with limited memory capacity. This setup reflects typical deployment constraints such as multi-tenant execution and edge-based inference nodes. Given these conditions, our experimental comparisons are conducted under controlled assumptions to highlight the performance characteristics of different deployment strategies, rather than to fully replicate production environments. The three deployment strategies are explained as follows:
\begin{itemize}

	\item \textbf{Cloud-only}: all inference tasks are executed in the cloud. This setting includes both a naive baseline that recomputes the system prompt for each query without cache reuse, and an optimized vLLM-based variant that precomputes and caches system prompt KV to accelerate batched inference~\cite{relayattention},~\cite{promptcache}.

	\item \textbf{Edge-only}:  all computations are performed entirely at the edge, without any collaborative inference or acceleration mechanisms. This serves as a baseline to evaluate the inference performance of edge devices under non-collaborative conditions.
	
	\item \textbf{Cloud-Edge} (ours): context processing and user prompt inference are performed separately on cloud and edge models. By jointly retrieving system prompt KV caches from both cloud and edge, the edge model is enabled to efficiently handle multiple user requests in parallel, supporting fast batched inference.
\end{itemize}
\begin{table}[t]
	\renewcommand{\arraystretch}{1.5}
	\begin{center}
		\caption{Specifications of GPU Used in Our Experiments.}
		\label{tab1}
		\begin{tabular}{ c|c|c|c}
			\hline
			\textbf{Symbols} &\textbf{Memory} & \textbf{Mem.Band} & \textbf{FP16 Peak F.}\\
			NVIDIA A800 80GB PCIe & 80 GB & 2030 GB/s & 77.9 TFLOPs \\		
			\hline
		\end{tabular}
	\end{center}
\end{table}
The primary performance evaluation metrics include: (1) End-to-end transmission time: the total time required to complete the entire inference task, including the cloud-based LLM computation time, communication transmission time, and the edge-based SLM computation time. (2) User-perceived latency, defined as the average time from request initiation to receipt of the complete response, serving as a measure of actual user experience;  (3) Normalized latency (ms/token), representing the average inference time per generated token; (4) Time to First Token (TTFT), measuring the time elapsed from request submission to the generation of the first token; and (5) Generation quality, evaluated by a weighted combination of BERTScore and ChatGPT-based scoring, where higher values indicate better semantic alignment between generated and reference texts. All experiments are repeated five times and averaged across batches to obtain stable single-inference performance measurements.

\renewcommand{\arraystretch}{3}
\begin{table*}[tbp]
	\centering
	\caption{Evaluation of Inference Efficiency and Resource Overhead under Different Deployment Strategies}
	\Huge
	\resizebox{\textwidth}{!}{
		\begin{tabular}{c|c|c|c|c|c|c|c|c|c}
			\hline
			\multicolumn{1}{c|}{\textbf{Dataset}}  & \multicolumn{2}{c|}{\textbf{Deployment stragety}}  &\multicolumn{1}{c|}{\textbf{TTFT(ms)}} & \multicolumn{1}{c|}{\textbf{Total Time Cost (s)}} & \multicolumn{1}{c|}{\textbf{Base Memory Usage (MB)}} & \multicolumn{1}{c|}{\textbf{Inference Memory (MB)}} & 
			\multicolumn{1}{c|}{\textbf{User Data Upload Ratio(\%)}} &    \multicolumn{1}{c|}{\textbf{Transmitted Data Size (MB)}} & \multicolumn{1}{c}{\textbf{Score}} \\ 
			\hline
			\multicolumn{1}{c|}{\multirow{4}{*}{xsum}} 
			&{\multirow{2}{*}{Cloud-only}} 
			& Naïve-cloud     & 652.12    &   519.63   &   12700.03  &  8.202     &   100      &      0.33       &  0.9025  \\ \cline{3-10}	
			& \multicolumn{1}{c|}{}                           
			& vLLM-ra &  304.25   &   78.72    &  13781.41    &  44075.00    &     100    &    0.33      & 0.896  \\ \cline{2-10}	
			& \multicolumn{1}{l|}{Edge-only}                 
			& Naïve-edge    & 311.592    &   384.34    &  2509.61       &      8.13   &    0   &  0   &  0.7009   \\ \cline{2-10}
			& \multicolumn{1}{l|}{Cloud-Edge}                  
			& CE-LSLM    & 210.24    &  68.58      &  2941.86   &   2756       &     0    &      0       & 0.877\\\cline{3-10}
			\hline
			\multirow{4}{*}{mmlu}
			& {\multirow{2}{*}{Cloud-only}} 
			& Naïve-cloud   &   412.99    & 199.11    &   12700.03   &   8.13    &      100   &     0.31     & 0.891  \\ \cline{3-10}
			& \multicolumn{1}{c|}{}    %
			& vLLM-ra &  335.59    &  26.01    &   13781.41   &   55488.00   &     100    &   0.31      & 0.873 \\ \cline{2-10}
			& \multicolumn{1}{l|}{Edge-only}                 
			& Naïve-edge   &  61.27   &   148.34    &     2509.61     & 8.13        &    0   &  0   & 0.72    \\ \cline{2-10}
			& \multicolumn{1}{l|}{Cloud-Edge}                  
			& CE-LSLM    &   170.03  &   11.00     &  2941.86  &    2756 &     0    &     0        &  0.854 \\ \cline{3-10}
			\hline
	\end{tabular}}
	
	\label{tab2}
\end{table*}

\subsection{Performance in Static Request Scenarios}

We selected 100 samples from each of the XSum ~\cite{xsum} and MMLU~\cite{hendrycks} datasets and modified them accordingly to suit the task scenarios within our request-response framework. Specifically, XSum article texts were used as contextual inputs, and task-specific queries were generated based on different topics. For MMLU, background information was added to enrich the prompt, and the original multiple-choice format was reformulated into open-ended question answering. The adapted inputs range from approximately 300 to 500 tokens in length and reflect medium-complexity inference and comprehension scenarios.

In the Cloud-only strategy, both the system prompt and user queries are required to be fully transmitted to the cloud for inference. Although the computational overhead associated with repeated processing of system prompts can be mitigated using caching mechanisms such as Relay Attention~\cite{relayattention}, all requests must still be centrally processed on the cloud. This leads to increased communication overhead and raises potential privacy concerns due to the transmission of sensitive user data. 

In contrast, under the proposed CE-LSLM strategy, sensitive user content is entirely processed on the edge device, with only deep-layer KV caches containing general knowledge being fetched from the cloud. For multi-batch inference tasks, the KV cache can be transmitted once to the edge and locally reused, thereby significantly reducing communication latency. In addition, model computation and cache loading are executed in parallel, eliminating extra waiting time. As shown in Table 2, CE-LSLM demonstrates a significant advantage in terms of Time to First Token (TTFT). In terms of generation quality, the BERTScore results are comparable to those of the Cloud-only strategy, indicating that the KV cache sharing mechanism enhances inference efficiency without compromising semantic consistency or output fidelity.

Under scenarios with concurrent requests or long-context inputs, the storage of intermediate states can significantly consume GPU memory. As shown in the table~\ref{tab2}, the peak memory requirement for vLLM-ra reaches up to $44GB$, which clearly exceeds the resource capacity of most edge devices. To improve request-handling capability, the Naive Edge-only strategy typically resorts to truncating the input context, thereby weakening semantic guidance and contextual grounding, which ultimately degrades the quality of the generated text.  Although edge models can deliver low response latency for lightweight tasks, the absence of parallel scheduling and cache reuse mechanisms leads to substantial inference delays when handling complex tasks. By comparison, CE-LSLM effectively mitigates the above resource bottlenecks through cache reuse strategies, enhancing both generation quality and the responsiveness of edge deployments.

The CE-COLLM framework incorporates an early exit mechanism, enabling partial inference to be completed at the edge without requiring cloud processing~\cite{CE-CoLLM}. However, this mechanism relies on a predefined confidence threshold if the confidence score of the generated output is low, the intermediate results will be transmitted to the cloud for further computation. Additionally, a stable network connection is required to prevent increased inference latency or task failure.

\begin{figure*}[t] 
	\centering

	\begin{subfigure}[b]{0.32\textwidth}
		\centering
		\includegraphics[width=2.3in,height=1.5in]{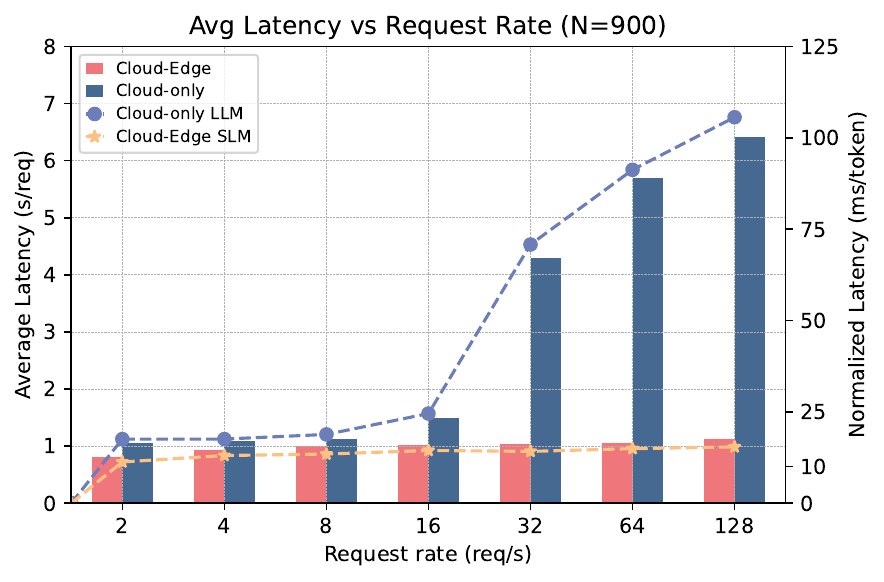}
		\caption{System prompt length=64}
			\label{fig8:a}
	\end{subfigure}%
	\hfil
	\begin{subfigure}[b]{0.32\textwidth}
		\centering
		\includegraphics[width=2.3in,height=1.5in]{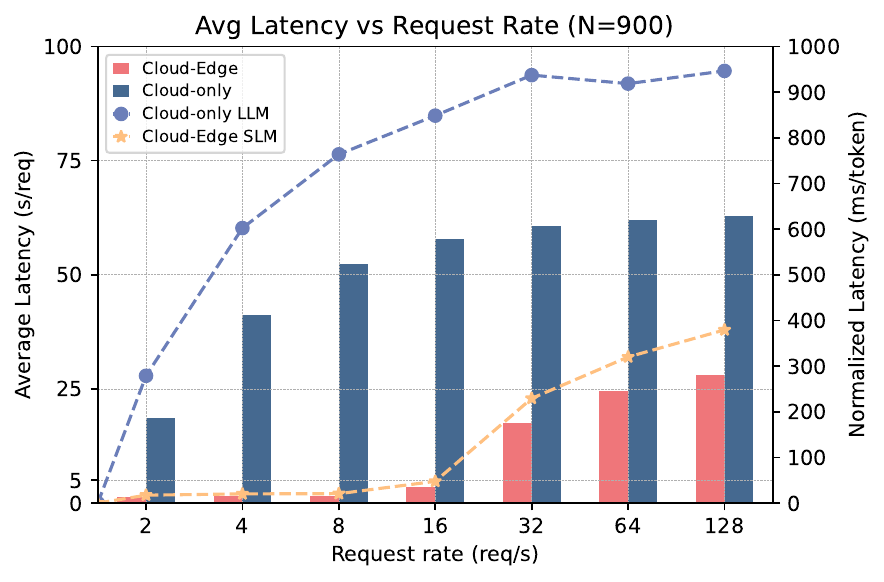}
		\caption{System prompt length=512}
		\label{fig8:b}
	\end{subfigure}
	\hfil
	\begin{subfigure}[b]{0.32\textwidth}
		\centering
		\includegraphics[width=2.3in,height=1.5in]{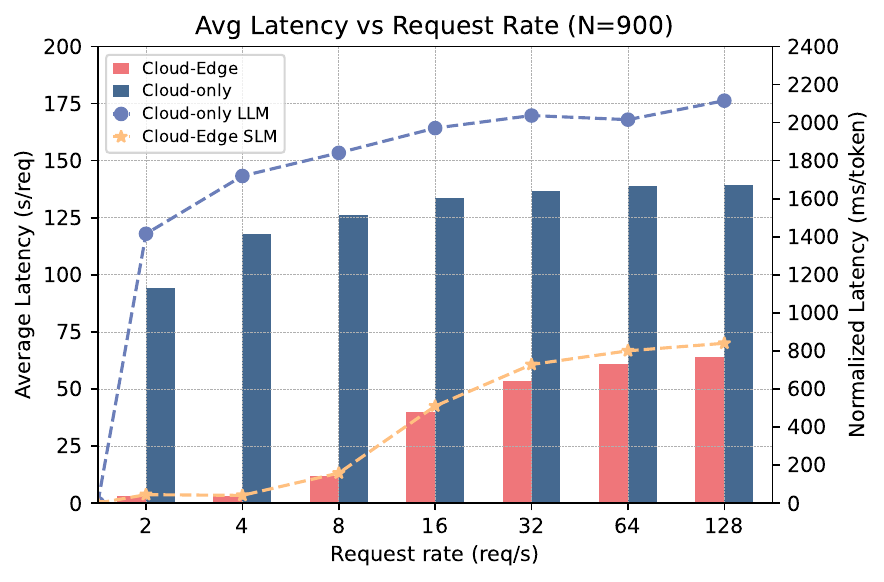}
		\caption{System prompt length=1024}
		\label{fig8:c}
	\end{subfigure}
	\hfil
	\begin{subfigure}[b]{0.32\textwidth}
	\centering
	\includegraphics[width=2.3in,height=1.5in]{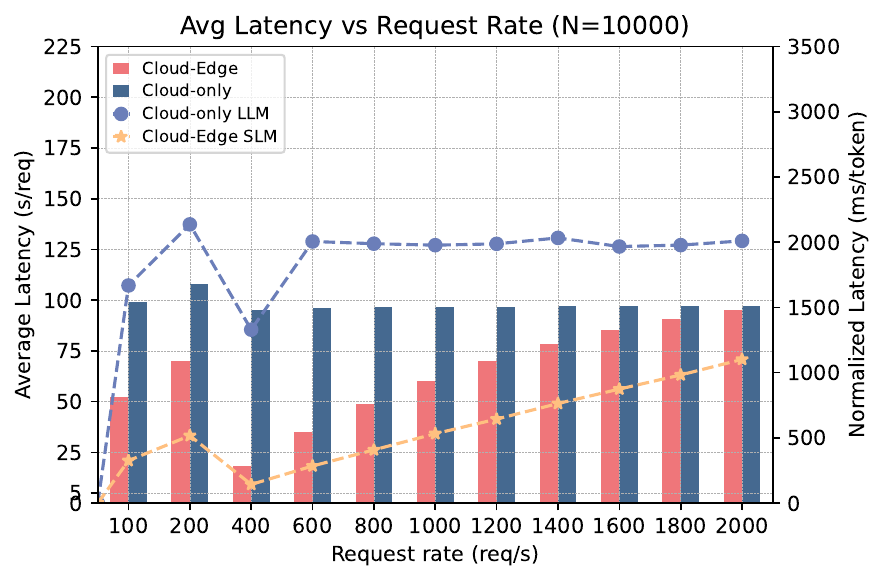}
	\caption{System prompt length=64}
	\label{fig8:d}
	\end{subfigure}%
	\hfil
	\begin{subfigure}[b]{0.32\textwidth}
	\centering
	\includegraphics[width=2.3in,height=1.5in]{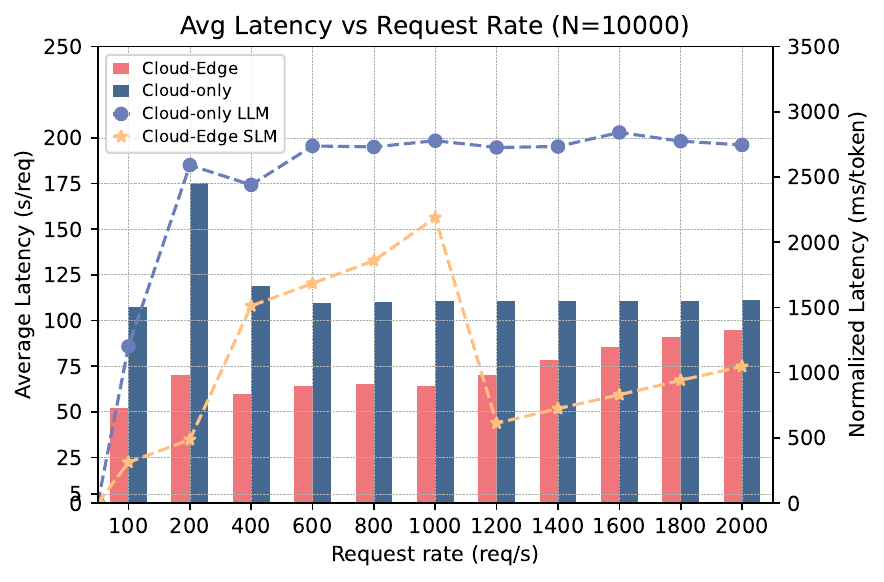}
	\caption{System prompt length=512}
	\label{fig8:e}
	\end{subfigure}
	\hfil
	\begin{subfigure}[b]{0.32\textwidth}
	\centering
	\includegraphics[width=2.3in,height=1.5in]{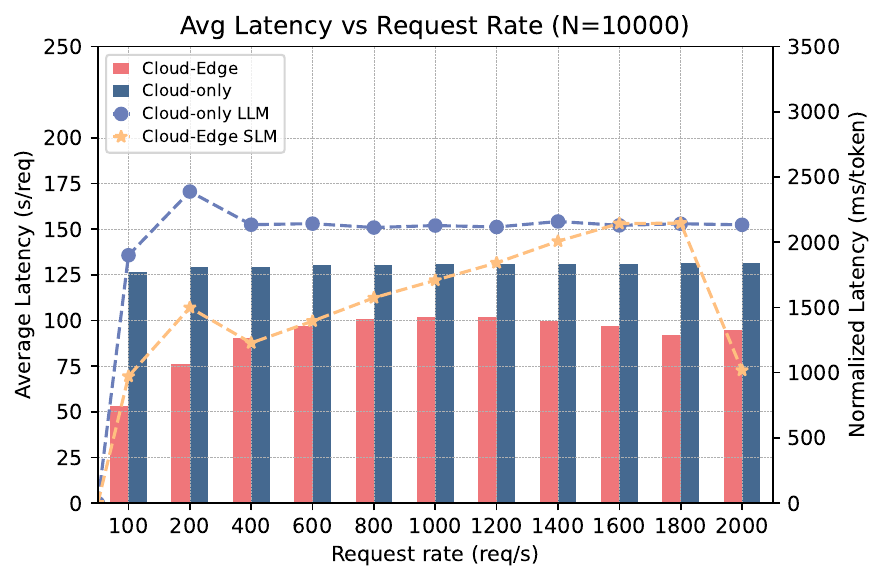}
	\caption{System prompt length=1024}
	\label{fig8:f}
	\end{subfigure}
	
	\caption{Specifically, subfigures (a)–(c) present the latency results under resource-constrained conditions, where GPU resources are concurrently shared among multiple users. In contrast, subfigures (d)–(f) illustrate latency outcomes in resource-sufficient settings. The left y-axis denotes the average response latency, while the right y-axis indicates the normalized latency.}
	\label{fig8}
\end{figure*}

\subsection{System Stability under High Request Rates}

In this experiment, the ShareGPT$\_$V3 dataset was used to simulate a high-concurrency request environment, aiming to evaluate system performance in terms of response latency and normalized latency under varying load intensities. A time-window-based scheduling strategy was employed to enhance controllability and measurement accuracy. LLM was configured to concurrently serve inference requests from five SLMs, covering both resource-constrained and resource-sufficient scenarios. Different prefix lengths were considered, and the maximum batch sizes supported by the memory capacity of both cloud and edge models were assigned to reflect memory-induced performance bottlenecks. It should be noted that the Naive Cloud-only and Naive Edge-only configurations were excluded from the comparison, as they fail to support sustained batch processing under resource constraints and frequently trigger memory overflows. The Naive Edge setup additionally suffers from unstable generation quality.

Fig. \ref{fig8} presents the inference latency results for cloud-based inference using vLLM and the proposed CE-LSLM under sustained high-concurrency request scenarios. Due to the deployment of SLMs at the edge, CE-LSLM exhibits real-time responsiveness. In lightweight input scenarios, CE-LSLM maintains an average inference latency of approximately 1 second across varying request rates, demonstrating high stability. In contrast, the LLM executes a greater number of attention operations per layer, incurring higher overhead in memory access and activation cache usage, thereby increasing per-token inference latency. As a result, when the request rate reaches 32 $req\/s$, the average latency of the Cloud-only configuration surges beyond 4 seconds—over four times the baseline latency—indicating evident queuing delays and performance degradation under moderate concurrency loads.

When the prefix length increases from 512 to 1024, as shown in Fig.~\ref{fig8}(b)-(c), systems operating under the compute-only execution mode experience significant nonlinear increases in response latency under high request rates. This behavior indicates that the system bottleneck gradually shifts from being communication-bound to computation-bound due to the inability to decouple compute and memory resources. In contrast, the latency growth of CE-LSLM remains relatively stable, demonstrating greater robustness. This advantage is primarily attributed to the reuse of KV caches and the parallel coordination of model computation and memory loading, which together alleviate memory access bottlenecks and significantly improve inference stability and concurrency handling capability.

When GPU resources are sufficient, as shown in Fig.~\ref{fig8}(d)-(f), both architectures gradually exhibit saturation behavior as the request rate increases significantly. However, CE-LSLM consistently achieves lower average latency compared to the Cloud-only configuration, indicating stronger delay control and greater system robustness under concurrent workloads and resource stress. In addition, when the prefix length exceeds 512, both average latency and normalized latency begin to show noticeable fluctuations. This behavior is attributed to the fact that small models are often used for short texts or lightweight tasks, where the shorter output sequences make per-token latency metrics more sensitive to minor temporal variations.

Overall, even under saturation conditions, the CE-LSLM architecture maintained effective task scheduling, edge cooperation, and caching mechanisms, without experiencing performance degradation due to increased network load. This reflects strong controllability of concurrent latency. By offloading a portion of the computation to the edge, the cloud-side burden was significantly reduced, enabling the system to support a larger number of edge-deployed SLM instances. In terms of throughput, at moderate request rates (e.g., 64 req/s), the CE-LSLM configuration achieved approximately 4–6× higher throughput compared to the Cloud-only baseline, thereby validating the effectiveness and rationality of edge-side role assignment in the collaborative inference architecture.

\section{Conclusion}
\noindent

In the process of extending generative AI services to wireless edge networks, limited memory and computational capacity have emerged as critical bottlenecks in the deployment of LLM. To address this challenge, a collaborative inference architecture was proposed, integrating LLM with SLMs. A novel service mechanism was explored in the context of 6G networks, aiming to jointly optimize generative processing and communication. The proposed architecture incorporates hierarchical cache sharing and a pipelined compute-load parallelism scheme, enabling coordinated optimization between inference execution and wireless communication resources. Furthermore, in scenarios where the cloud is unavailable or network connections are temporarily disrupted, partial inference tasks can still be executed by edge models through shared cache reuse, exhibiting an initial form of distributed autonomy. Experimental results demonstrated that the proposed system significantly outperformed the conventional cloud-only baseline in terms of edge-side response latency and throughput, while maintaining comparable generation quality. These results further highlighted the system’s potential in supporting privacy preservation and scalability in multi-tenant edge deployment scenarios.

In addition, this work has theoretically and empirically demonstrated the feasibility of enabling collaborative inference through cache sharing across both cloud-edge and edge-edge model deployments. Future work will focus on exploring efficient state synchronization and scheduling mechanisms among edge-side models to support more flexible and resilient distributed inference strategies.

\bibliographystyle{IEEEtran}
\bibliography{ce_lslm}

\begin{thebibliography}{10}
\providecommand{\url}[1]{#1}
\csname url@samestyle\endcsname
\providecommand{\newblock}{\relax}
\providecommand{\bibinfo}[2]{#2}
\providecommand{\BIBentrySTDinterwordspacing}{\spaceskip=0pt\relax}
\providecommand{\BIBentryALTinterwordstretchfactor}{4}
\providecommand{\BIBentryALTinterwordspacing}{\spaceskip=\fontdimen2\font plus
\BIBentryALTinterwordstretchfactor\fontdimen3\font minus
  \fontdimen4\font\relax}
\providecommand{\BIBforeignlanguage}[2]{{%
\expandafter\ifx\csname l@#1\endcsname\relax
\typeout{** WARNING: IEEEtran.bst: No hyphenation pattern has been}%
\typeout{** loaded for the language `#1'. Using the pattern for}%
\typeout{** the default language instead.}%
\else
\language=\csname l@#1\endcsname
\fi
#2}}
\providecommand{\BIBdecl}{\relax}
\BIBdecl

\bibitem{ref6G}
M.~Z. Chowdhury, M.~Shahjalal, S.~Ahmed, and Y.~M. Jang, ``6g wireless
  communication systems: Applications, requirements, technologies, challenges,
  and research directions,'' \emph{IEEE Open Journal of the Communications
  Society}, vol.~1, pp. 957--975, 2020.

\bibitem{white}
N.~Rajatheva, I.~Atzeni, E.~Bjornson, A.~Bourdoux, S.~Buzzi, J.-B. Dore,
  S.~Erkucuk, M.~Fuentes, K.~Guan, Y.~Hu \emph{et~al.}, ``White paper on
  broadband connectivity in 6g,'' \emph{arXiv preprint arXiv:2004.14247}, 2020.

\bibitem{AI-native}
W.~Wu, C.~Zhou, M.~Li, H.~Wu, H.~Zhou, N.~Zhang, X.~S. Shen, and W.~Zhuang,
  ``Ai-native network slicing for 6g networks,'' \emph{IEEE Wireless
  Communications}, vol.~29, no.~1, pp. 96--103, 2022.

\bibitem{6G}
X.~You, C.-X. Wang, J.~Huang, X.~Gao, Z.~Zhang, M.~Wang, Y.~Huang, C.~Zhang,
  Y.~Jiang, J.~Wang \emph{et~al.}, ``Towards 6g wireless communication
  networks: Vision, enabling technologies, and new paradigm shifts,''
  \emph{Science China Information Sciences}, vol.~64, pp. 1--74, 2021.

\bibitem{optimizing}
S.~Hisaharo, Y.~Nishimura, and A.~Takahashi, ``Optimizing llm inference
  clusters for enhanced performance and energy efficiency,'' \emph{Authorea
  Preprints}, 2024.

\bibitem{performance}
J.~Kundu, W.~Guo, A.~BanaGozar, U.~De~Alwis, S.~Sengupta, P.~Gupta, and
  A.~Mallik, ``Performance modeling and workload analysis of distributed large
  language model training and inference,'' in \emph{2024 IEEE International
  Symposium on Workload Characterization (IISWC)}.\hskip 1em plus 0.5em minus
  0.4em\relax IEEE, 2024, pp. 57--67.

\bibitem{deepspeed}
J.~Rasley, S.~Rajbhandari, O.~Ruwase, and Y.~He, ``Deepspeed: System
  optimizations enable training deep learning models with over 100 billion
  parameters,'' in \emph{Proceedings of the 26th ACM SIGKDD International
  Conference on Knowledge Discovery \& Data Mining}, 2020, pp. 3505--3506.

\bibitem{reducing}
V.~A. Korthikanti, J.~Casper, S.~Lym, L.~McAfee, M.~Andersch, M.~Shoeybi, and
  B.~Catanzaro, ``Reducing activation recomputation in large transformer
  models,'' \emph{Proceedings of Machine Learning and Systems}, vol.~5, pp.
  341--353, 2023.

\bibitem{gpuclusters}
D.~Narayanan, M.~Shoeybi, J.~Casper, P.~LeGresley, M.~Patwary, V.~Korthikanti,
  D.~Vainbrand, P.~Kashinkunti, J.~Bernauer, B.~Catanzaro \emph{et~al.},
  ``Efficient large-scale language model training on gpu clusters using
  megatron-lm,'' in \emph{Proceedings of the International Conference for High
  Performance Computing, Networking, Storage and Analysis}, 2021, pp. 1--15.

\bibitem{pagedattention}
W.~Kwon, Z.~Li, S.~Zhuang, Y.~Sheng, L.~Zheng, C.~H. Yu, J.~Gonzalez, H.~Zhang,
  and I.~Stoica, ``Efficient memory management for large language model serving
  with pagedattention,'' in \emph{Proceedings of the 29th Symposium on
  Operating Systems Principles}, 2023, pp. 611--626.

\bibitem{relayattention}
L.~Zhu, X.~Wang, W.~Zhang, and R.~W. Lau, ``Relayattention for efficient large
  language model serving with long system prompts,'' \emph{arXiv preprint
  arXiv:2402.14808}, 2024.

\bibitem{promptcache}
I.~Gim, G.~Chen, S.-s. Lee, N.~Sarda, A.~Khandelwal, and L.~Zhong, ``Prompt
  cache: Modular attention reuse for low-latency inference,'' \emph{Proceedings
  of Machine Learning and Systems}, vol.~6, pp. 325--338, 2024.

\bibitem{model}
Y.~Jiang, S.~Wang, V.~Valls, B.~J. Ko, W.-H. Lee, K.~K. Leung, and
  L.~Tassiulas, ``Model pruning enables efficient federated learning on edge
  devices,'' \emph{IEEE Transactions on Neural Networks and Learning Systems},
  vol.~34, no.~12, pp. 10\,374--10\,386, 2022.

\bibitem{edgellm}
\BIBentryALTinterwordspacing
Z.~Yu, Z.~Wang, Y.~Li, H.~You, R.~Gao, X.~Zhou, S.~R. Bommu, Y.~K. Zhao, and
  Y.~C. Lin, ``Edge-llm: Enabling efficient large language model adaptation on
  edge devices via layerwise unified compression and adaptive layer tuning and
  voting,'' 2024. [Online]. Available: \url{https://arxiv.org/abs/2406.15758}
\BIBentrySTDinterwordspacing

\bibitem{AIGX}
Y.~Liu, H.~Du, D.~Niyato, J.~Kang, S.~Cui, X.~Shen, and P.~Zhang, ``Optimizing
  mobile-edge ai-generated everything (aigx) services by prompt engineering:
  Fundamental, framework, and case study,'' \emph{IEEE Network}, 2023.

\bibitem{CE-CoLLM}
H.~Jin and Y.~Wu, ``Ce-collm: Efficient and adaptive large language models
  through cloud-edge collaboration,'' \emph{arXiv preprint arXiv:2411.02829},
  2024.

\bibitem{infinite}
B.~Lin, C.~Zhang, T.~Peng, H.~Zhao, W.~Xiao, M.~Sun, A.~Liu, Z.~Zhang, L.~Li,
  X.~Qiu \emph{et~al.}, ``Infinite-llm: Efficient llm service for long context
  with distattention and distributed kvcache,'' \emph{arXiv preprint
  arXiv:2401.02669}, 2024.

\bibitem{zero}
S.~Rajbhandari, J.~Rasley, O.~Ruwase, and Y.~He, ``Zero: Memory optimizations
  toward training trillion parameter models,'' in \emph{SC20: International
  Conference for High Performance Computing, Networking, Storage and
  Analysis}.\hskip 1em plus 0.5em minus 0.4em\relax IEEE, 2020, pp. 1--16.

\bibitem{llmpruner}
\BIBentryALTinterwordspacing
X.~Ma, G.~Fang, and X.~Wang, ``Llm-pruner: On the structural pruning of large
  language models,'' 2023. [Online]. Available:
  \url{https://arxiv.org/abs/2305.11627}
\BIBentrySTDinterwordspacing

\bibitem{sparse}
\BIBentryALTinterwordspacing
E.~Frantar and D.~Alistarh, ``Sparsegpt: Massive language models can be
  accurately pruned in one-shot,'' 2023. [Online]. Available:
  \url{https://arxiv.org/abs/2301.00774}
\BIBentrySTDinterwordspacing

\bibitem{cachegen}
Y.~Liu, H.~Li, Y.~Cheng, S.~Ray, Y.~Huang, Q.~Zhang, K.~Du, J.~Yao, S.~Lu,
  G.~Ananthanarayanan \emph{et~al.}, ``Cachegen: Kv cache compression and
  streaming for fast large language model serving,'' in \emph{Proceedings of
  the ACM SIGCOMM 2024 Conference}, 2024, pp. 38--56.

\bibitem{kvsharer}
\BIBentryALTinterwordspacing
Y.~Yang, Z.~Cao, Q.~Chen, L.~Qin, D.~Yang, H.~Zhao, and Z.~Chen, ``Kvsharer:
  Efficient inference via layer-wise dissimilar kv cache sharing,'' 2024.
  [Online]. Available: \url{https://arxiv.org/abs/2410.18517}
\BIBentrySTDinterwordspacing

\bibitem{adakv}
\BIBentryALTinterwordspacing
Y.~Feng, J.~Lv, Y.~Cao, X.~Xie, and S.~K. Zhou, ``Ada-kv: Optimizing kv cache
  eviction by adaptive budget allocation for efficient llm inference,'' 2024.
  [Online]. Available: \url{https://arxiv.org/abs/2407.11550}
\BIBentrySTDinterwordspacing

\bibitem{discard}
S.~Ge, Y.~Zhang, L.~Liu, M.~Zhang, J.~Han, and J.~Gao, ``Model tells you what
  to discard: Adaptive kv cache compression for llms,'' in \emph{Workshop on
  Advancing Neural Network Training: Computational Efficiency, Scalability, and
  Resource Optimization (WANT@ NeurIPS 2023)}.

\bibitem{netgpt}
Y.~Chen, R.~Li, Z.~Zhao, C.~Peng, J.~Wu, E.~Hossain, and H.~Zhang, ``Netgpt: An
  ai-native network architecture for provisioning beyond personalized
  generative services,'' \emph{IEEE Network}, 2024.

\bibitem{computeorload}
S.~Jin, X.~Liu, Q.~Zhang, and Z.~M. Mao, ``Compute or load kv cache? why not
  both?'' \emph{arXiv preprint arXiv:2410.03065}, 2024.

\bibitem{osca}
Y.~Zhang, P.~Huang, K.~Zhou, H.~Wang, J.~Hu, Y.~Ji, and B.~Cheng,
  ``$\{$OSCA$\}$: An $\{$Online-Model$\}$ based cache allocation scheme in
  cloud block storage systems,'' in \emph{2020 USENIX Annual Technical
  Conference (USENIX ATC 20)}, 2020, pp. 785--798.

\bibitem{similarity}
\BIBentryALTinterwordspacing
M.~Klabunde, T.~Schumacher, M.~Strohmaier, and F.~Lemmerich, ``Similarity of
  neural network models: A survey of functional and representational
  measures,'' \emph{ACM Computing Surveys}, Apr. 2025. [Online]. Available:
  \url{http://dx.doi.org/10.1145/3728458}
\BIBentrySTDinterwordspacing

\bibitem{DBLP}
\BIBentryALTinterwordspacing
S.~Kornblith, M.~Norouzi, H.~Lee, and G.~E. Hinton, ``Similarity of neural
  network representations revisited,'' \emph{CoRR}, vol. abs/1905.00414, 2019.
  [Online]. Available: \url{http://arxiv.org/abs/1905.00414}
\BIBentrySTDinterwordspacing

\bibitem{think}
Y.~Xu, Z.~Jie, H.~Dong, L.~Wang, X.~Lu, A.~Zhou, A.~Saha, C.~Xiong, and
  D.~Sahoo, ``Think: Thinner key cache by query-driven pruning,'' \emph{arXiv
  preprint arXiv:2407.21018}, 2024.

\bibitem{survey}
W.~X. Zhao, K.~Zhou, J.~Li, T.~Tang, X.~Wang, Y.~Hou, Y.~Min, B.~Zhang,
  J.~Zhang, Z.~Dong \emph{et~al.}, ``A survey of large language models,''
  \emph{arXiv preprint arXiv:2303.18223}, vol.~1, no.~2, 2023.

\bibitem{xsum}
S.~Narayan, S.~B. Cohen, and M.~Lapata, ``Don't give me the details, just the
  summary! topic-aware convolutional neural networks for extreme
  summarization,'' \emph{ArXiv}, vol. abs/1808.08745, 2018.

\bibitem{hendrycks}
D.~Hendrycks, C.~Burns, S.~Basart, A.~Critch, J.~Li, D.~Song, and
  J.~Steinhardt, ``Aligning ai with shared human values,'' \emph{Proceedings of
  the International Conference on Learning Representations (ICLR)}, 2021.

\end{thebibliography}

\clearpage 
\appendix

\section{Derivation of the Attention Formula}
\label{appendix:derivation}

In this appendix, we provide a detailed mathematical derivation of the prunable distance length for attention heads in ALiBi-based models.

\subsection{Invariance Properties of CKA}
\label{appendix:CKA}

The Centered Kernel Alignment (CKA) similarity metric exhibits several desirable invariance properties that make it particularly suitable for comparing neural network representations~\cite{DBLP}:

\subsubsection{Scale invariance}
\
We consider the representation matrix of each layer denoted as $O_{e}\in \mathbb{R}^{n\times d}$, where nn is the sequence length and dd is the hidden dimension. Let 
\begin{equation}
	O_{c}^{\left ( l \right )}=\alpha O_{e}^{\left ( l \right )}, \ \ \alpha \in \mathbb{R} 
\end{equation}

This indicates that all sample representations are scaled by a shared constant factor $\alpha$, which is commonly observed in operations such as LayerNorm and residual connections. 

Then,  the kernel matrices can be computed as follows:
\begin{equation} \label{scale}
	\begin{aligned}
		S_{c}=O_{c}O_{c}^{T}=\alpha ^{2}O_{e}O_{e}^{T} =\alpha ^{2}S_{e}
	\end{aligned}
\end{equation}

The following formulation can be derived by combining Eqs.~\ref{HSIC} and ~\ref{scale} :
\begin{equation} \label{scale_kernel}
	\begin{aligned}
	HSIC\left ( S_{e},S_{c}\right ) &=\frac{1}{\left ( N-1 \right )^{2} } tr\left ( H_{N}S_{e}H_{N}\alpha^{2} S_{e} \right ) \\
	&=\alpha ^{2}\cdot HSIC\left ( S_{e},S_{e}\right ) 
	\end{aligned}
\end{equation}

By substituting the above formulation into the normalization Eqs.~\ref{CKA}, we have
\begin{equation}
	\begin{aligned}
		m_{CKA} \left (  S_{e},S_{c}\right ) &=\frac{\alpha ^{2}\cdot HSIC\left ( S_{e},S_{e}\right ) }{\sqrt{ HSIC\left ( S_{e},S_{e}\right )\cdot \alpha^{4} \cdot HSIC\left ( S_{e},S_{e} \right )  } } \\
		&=1
	\end{aligned}
\end{equation}

Hence, we have 
\begin{equation}
	m_{CKA} \left (  O_{e},\alpha O_{e}\right ) = m_{CKA} \left (  O_{e},O_{c}\right ) 
\end{equation}

\subsubsection{Orthogonal invariance}
\
Let the layer representation matrix be $O_{e} \in \mathbb{R}^{n \times d}$. After applying an orthogonal transformation, it becomes:
\begin{equation}\label{Oc}
	O_{c}=O_{e}Q
\end{equation}
where  $Q^{T}Q=I$.

Subsequently, the kernel matrix can be expressed as:
\begin{equation}
	\begin{aligned}
		S_{e}=O_{e}O_{e}^{T} 
	\end{aligned}
\end{equation}

Since orthogonal transformations do not alter the inner product structure between samples, it follows that:
\begin{equation}
	\begin{aligned}
		S_{c}=O_{c}O_{c}^{T}=(O_{e}Q)(O_{e}Q)^{T}=O_{e}QQ^{T}O_{e}^{T}=S
	\end{aligned}
\end{equation}

Finally, it can be verified that
\begin{equation}\label{Ssi}
	\begin{aligned}
	HSIC\left(S_{e}, S_{c}\right)&=HSIC\left(S_{e}, S_{e}\right) \\
	m_{CKA}\left(S_{e}, S_{c}\right)&=1
	\end{aligned}	
\end{equation}

\subsubsection{Permutation invariance}
\
Given a layer representation matrix $O_{e}\in \mathbb{R}^{n\times d}$, after a permutation of the feature dimensions, that is,
\begin{equation}
	O_{c}=O_{e}P
\end{equation}
where $P$ is a permutation matrix.

Accordingly, the kernel matrix can be expressed as
\begin{equation}
	S_{e}=O_{e}O_{e}^{T}
\end{equation}

The permutation matrix satisfies $P^{\top} = P^{-1}$ and $PP^{\top} = I$, thus the inner product remains unchanged, and can be expressed as
\begin{equation}
  S_{c}=O_{c}O_{c}^{T}=(O_{e}P)(O_{e}P)^{T}=O_{e}PP^{T}O_{e}^{T}=S
\end{equation}

Thus, CKA can be defined as
Taking the logarithm of both sides of the above equation yields:
\begin{equation}\label{Ssi}
		m_{CKA}\left(S_{e}, S_{c}\right)=m_{CKA}\left(S_{e}, S_{e}\right)=1
\end{equation}

\end{document}